\documentclass[12pt]{article}
\newcommand{\bea}{\begin{eqnarray}}
\newcommand{\eea}{\end{eqnarray}}
\newcommand{\be}{\begin{equation}}
\newcommand{\ee}{\end{equation}}

\def\be{\begin{eqnarray}}
\def\ee{\end{eqnarray}}
\def\bd{\begin{displaymath}}
\def\ed{\end{displaymath}}

\def\etal{{\em et al. }}

\def\NP{Nucl. Phys. }
\def\PR{Phys. Rev. }
\def\PRL{Phys. Rev. Lett. }
\def\PL{Phys. Lett. }
\def\jpg{J. Phys. G: Nucl. Part. Phys. }

\begin{document}

\title{Microscopic calculation of half lives of spherical proton emitters}

\author{Madhubrata Bhattacharya
and G. Gangopadhyay\\
Department of Physics, University of Calcutta\\
92, Acharya Prafulla Chandra Road, Kolkata-700 009, India\\
email:ggphy@caluniv.ac.in}
\maketitle

\begin{abstract}

Half life values for proton radioactivity in nuclei have been calculated 
in the WKB approximation. The microscopic proton-nucleus potential
has been obtained by folding the densities of daughter nuclei with two
microscopic NN interactions, DDM3Y and JLM. 
The densities have been obtained in the Relativistic Mean Field
approach in the spherical approximation using the force FSU Gold.
No substantial modification of results has been observed if other common 
forces are employed. The calculated results for the decays from the
ground state or the low-lying excited states in almost all the nuclei
agree well with experimental measurements. Reasons for large deviations 
in a few cases have been discussed. Results in $^{109}$I and $^{112,113}$Cs 
show that the effect of deformation is small contrary to earlier calculations.
Predictions for possible proton radioactivity have been made in two nuclei, 
$^{93}$Ag and $^{97}$In.

\end{abstract}



\vskip 0.5cm
Proton drip line nuclei play a very important role in nuclear astrophysics,
namely in nova and supernova explosions, X-ray bursts associated with explosive 
hydrogen burning,  rapid proton capture processes, etc. These processes 
involve the low excitation energy  regime of the nucleus. The rapid proton 
capture process at low energy has its inverse in the proton radioactivity from 
the nuclear ground
state or low energy states. In both the cases, the important feature is the
Coulomb barrier which has to be overcome by the proton through tunneling. 
It is thus very important to study the proton radioactivity process.
 
Although the first example of proton radioactivity from  nuclei was observed 
in an isomeric state of $^{53}$Co in 1970, the next decay was observed only a 
decade later in $^{151}$Lu. Since then, with the improvement in experimental 
facilities, examples of proton 
radioactivity from ground states or low lying isomeric states has been found
in a number of nuclei, all between $Z=51$ and $Z=83$\cite{prdecay,nds}. 

Theoretical calculations have been employed to explain the observed
lifetimes of proton radioactivity. Most of the investigations employ the 
picture of a point proton tunneling through a potential barrier and calculate 
the half life in the WKB approximation. The potential that has been 
predominantly used is the Woods-Saxon form added to the Coulomb potential
\cite{Buck,Aberg,mfl}. Recently Basu {\em et al.} have 
constructed the proton-nucleus nuclear potential in the single folding model
and used it to calculate the decay probability in observed spherical proton 
emitters\cite{oth}. However, in all these calculations, the density of the 
daughter nucleus is taken from phenomenological models. 
Replacement of this density by one obtained from some
mean field calculation may be expected to reproduce the results more accurately.

                                                                                
Relativistic Mean Field (RMF) approach is now a standard tool in low energy
nuclear structure.
It has been able to explain different features
of stable and exotic nuclei like ground state binding energy, deformation,
radius, excited states, spin-orbit splitting, neutron halo, etc\cite{RMF1}.
It is well known that in nuclei far away from the
stability valley, the single particle level structure undergoes certain
changes in which the spin-orbit splitting plays an important role.
Being based on the Dirac Lagrangian density, RMF is particularly suited to 
investigate these nuclei because it naturally incorporates the
spin degrees of freedom. 
Relativistic Hartree Bogoliubov (RHB) calculations
have been used to predict the proton drip line in medium-heavy to superheavy
nuclei with considerable success\cite{rhb}.

There exist different variations of the Lagrangian density as
well as a number of different parameterizations. Recently, a new Lagrangian
density has been proposed\cite{prl} which involves
self-coupling of the vector-isoscalar meson as well as coupling between the
vector-isoscalar meson and the vector-isovector meson. The corresponding 
parameter set is 
called FSU Gold\cite{prl}. In this work, we have employed mainly this force 
for our calculation, though we have also checked our calculations using the
forces NL3\cite{NL3} and NLSH\cite{NLSH} in some cases.

In the conventional RMF+BCS approach for even-even nuclei, the Euler-Lagrange
equations are solved under the assumptions of classical meson
fields, time reversal symmetry, no-sea contribution, etc. Pairing is introduced
under the BCS approximation. Since accuracy of the nuclear density is very
important in our calculation, we have solved the equations in co-ordinate space.
The strength of the zero range pairing force is taken as 300 MeV-fm for both 
protons and neutrons. These values have been chosen to represent a good fit 
for the binding energy values in the daughter nuclei. For odd number of 
nucleons, the tagging approximation has been used to specify the level
occupied by the last odd nucleon of either type. We have observed that 
moderate variations of the pairing strength do not influence the life time 
to any great extent.

The effective NN interactions density dependent M3Y (DDM3Y)\cite{ddm3y1,ddm3y2} and 
that of Jeukenne, Lejeune and Mahaux (JLM)\cite{jlm} have been used to 
construct the proton nucleus potential. Both these interactions have been 
derived from nuclear matter calculation and have been applied in finite nuclei
with success.  

The DDM3Y interaction\cite{ddm3y1,ddm3y2} is obtained from a finite range 
energy independent M3Y interaction by adding a zero range energy dependent 
pseudopotential and introducing a density dependent factor. 
The density dependence may be chosen as exponential\cite{ddm3y1} or be
of the more physical form  $C(1-\beta\rho^{2/3})$\cite{ddm3y2}.
The constants were obtained from nuclear matter calculation\cite{ddm3y3}
as $C=2.07$ and $\beta=1.624$ fm$^2$. We have used this form in our 
calculation with the above parameters.
This 
interaction has been employed widely in the study of nucleon nucleus as well 
as nucleus nucleus scattering, calculation of proton radioactivity, etc.  

The JLM potential\cite{jlm} has been applied to finite nuclei in the Local 
Density Approximation. Finite range of the interaction has been incorporated by 
including a Gaussian form factor\cite{jlm1}. We have already applied the JLM 
potential to obtain the semimicroscopic optical model potential and to study 
elastic scattering in lighter nuclei with success\cite{CBe}. Here we have used 
global parameters for the interaction and the default normalizations from Bauge
{\em et al.}\cite{jlm1}. We have not come across any earlier work which uses 
the JLM interaction to calculate decay probabilities.

The microscopic nuclear potential has been obtained by folding the
DDM3Y or JLM interactions with the 
microscopic densities obtained in the RMF calculation. 
The Coulomb potential has been similarly obtained by folding the Coulomb
interaction with the microscopic proton density.  The total potential 
consists of the nuclear part, the Coulomb potential as well as the centrifugal
potential. We have not included the contribution of isovector component of the 
folded potential {\em i.e.} the  Lane potential. However, we expect its effect 
to be 
small.
In the case of JLM interaction, 
we choose only the real part of the potential. The half life  of the 
parent nucleus has been obtained from the probability of barrier
penetration in the WKB approximation. The assault frequency is obtained from
the zero-point vibration energy that, in turn, has been calculated from the
Q-values following the prescription of Poenaru {\em et al.}\cite{ev} for
cluster radioactivity extended for protons.
The details of the calculations can be obtained from other references,
{\em eg.} \cite{oth} and are not detailed here. Unlike some other  
works, we have not multiplied the proton nucleus potential by any 
normalization factor and have used the values that were obtained from 
nuclear matter calculations. 

The calculated binding energy values of the daughter nuclei for the 
decays studied are compared with experimental or estimated values in
Table \ref{be}. One can see that the results are reasonably good. 
In the bottom part of the table, the binding energy values are presented for 
a few other nuclei which are not known to be associated with proton 
radioactivity experimentally. We will discuss their significance later.

The results for the proton radioactivity half life calculation, in nuclei
where such decay has been observed from low energy states, are tabulated in 
Table \ref{life1} for odd mass parents and Table \ref{life2} for odd-odd ones
along with the experimental values and their errors. 
For comparison, we have also presented the results for the DDM3Y interaction
of Basu \etal\cite{oth}, who have used a simple phenomenological distribution 
for the densities. 
The experimental values are obtained from the compilation\cite{nds} by Sonzogni.
We have also presented the uncertainties in the calculated half life values
for the errors in the measured Q-values within parentheses.
In a few nuclei, the results for Ref. \cite{oth} were
obtained using slightly different experimental Q-values as indicated in the
tables. More recent Q-values from \cite{nds} are expected to modify their 
results slightly.
We would like to emphasize that in most of these cases, our results compare
favourably with that of \cite{oth}. 

We note that the result for $^{105}$Sb 
using the JLM interaction is substantially different from that using DDM3Y
interaction. This is possibly due to the very low Q-value of the decay.
In all other cases, where the
Q-values are much larger, close to or greater than 1 
MeV, 
the results for the two interactions are very close
to each other as well as to those of \cite{oth}.
It signifies that while in this energy range, the behaviours of DDM3Y and 
JLM are similar, at very low energy, they may be different.
We also note that 
a recent work\cite{Liu} failed to observe any proton radioactivity in 
$^{105}$Sb.
Thus it is not possible to 
comment on the relative merits of the two interactions.

One can see that the results are also in very good agreement with experimental 
values in most of the cases. In a number of decays,  the results do 
not match so well, {\em eg.} in $^{147}$Tm, $^{150}$Lu  or $^{156}$Ta and more 
prominently in $^{185}$Bi and in the decay from the excited state of $^{177}$Tl,
the last two results being off by an order of magnitude. 

In the case of $^{177}$Tl, the 
half life of the decay from the ground state is reproduced very well. We have 
assumed that all the parent states have one quasi-proton configuration. 
The excited state involved in the proton radioactivity in this nucleus is 
situated 0.8 MeV above the ground state. At this energy the state may have 
substantial contribution from three quasi-proton configurations thus hindering 
the decay. Such contributions may also play a role in some other cases where 
the agreement is poorer.

The hindrance in $^{185}$Bi may also be explained easily. Davids \etal
\cite{Bi185} observed proton radioactivity in  $^{185}$Bi from $1/2^+$ state 
of the parent to the ground state of the daughter. The parent state is 
expected to have the configuration $\pi(h_{9/2})^2(s_{1/2})^{-1}$ configuration and yet it decays to the ground state of $^{184}$Pb. Thus this decay is more 
complicated and Davids \etal suggested that it takes place through the small 
$2p-2h$ admixture in the ground state of the daughter\cite{Bi185}. The 
calculated admixture fraction of the $2p-2h$ state to the ground state comes 
out to be $\sim 0.08$. This value is consistent with calculation in 
$^{186,188}$Pb\cite{Pb} which gives values $\le 0.09$. The $2p-2h$ 0$^+$ state
in the daughter $^{184}$Pb is estimated at 0.6 MeV from systematics. 
Using this value, the half life for the decay to the ground state 
has been calculated. The branching ratio for the decay to this state 
is found out to be very small, of the order of 10$^{-5}$ only. So the
decay to the ground state is expected to dominate.

Systematics of light Bi isotopes indicate that the lowest state is the
one quasi-proton state $9/2^-$. Davids \etal also assumed that 
the observed  $1/2^+$ state involved in the proton radioactivity is an excited 
state. However as no proton or alpha radioactivity was observed which involves
the  $9/2^-$ state, it has been concluded that the $1/2^+$ state is the ground 
state\cite{nds}. It is possible to predict the half life of the
decay from the above mentioned $9/2^-$ state to the ground state of $^{184}$Pb
assuming the Q-value of the decay to be the same as that for the decay from the 
$1/2^+$ state. The half life of the decay (with $l=5$)  comes out to be 26 ms. 
This value must be treated as the lower limit in case the $9/2^-$ is the ground 
state.

We want to emphasize a very important result of our calculation.
It has often been suggested that the half life values for proton radioactivity
of $^{109}$I and $^{112,113}$Cs cannot be reproduced without the 
inclusion of deformation effects. It was pointed out that a deformation 
of the order of $\beta\sim 0.05-0.15$ is essential to reproduce this data
\cite{sjnp}.
For example, Maglione {\em et al.}\cite{mfl} have pointed out that none of the 
single particle states have reproduced the observed half life value
within an order of magnitude. The authors used the 
Woods Saxon potential to obtain the single particle states. They observed that
a deformation of $\beta\sim 0.15$ is essential for reproduction of the
experimental lifetime.
However, our calculation reproduces the data for $^{109}$I with considerable 
accuracy. In the deformed calculations, it is usually assumed that the
deformation of the parent and the daughter nuclei are identical. The daughter 
in this particular case is $^{108}$Te. Te nuclei are well known vibrational
nuclei with very small deformation. One possibility may be that it is the 
deficiency of the Woods Saxon potential far away from the stability valley 
which is responsible 
for the failure of the calculations. We also stress that our results are
nearly identical for both the NN interactions. The results for 
$^{113}$Cs and to some extent $^{112}$Cs are not reproduced so well and 
this shortcoming may be an effect of deformation. However, in none of them do
we have an order of magnitude disagreement between theory and experiment as
obtained in \cite{mfl}. 

To verify that our results  are not peculiar to the chosen Lagrangian density,
we have recalculated the results of three nuclei near the beginning, middle and 
the end of the mass region studied using the densities NL3\cite{NL3} and NLSH
\cite{NLSH}. All the calculations have used the
DDM3Y interaction. We present the results in Table \ref{rmf}. We see that the 
results are very close to each other, 
{\em i.e.} they are independent of the force chosen. This is
true even for $^{113}$Cs. The uncertainties in half life values due to the 
errors in measured Q-values are identical with those of our calculated values 
presented in Table \ref{life1} and are not shown.

We would also like to investigate the possibilities of observing proton 
radioactivity
in nearby nuclei and chosen  $^{97}$In, $^{93}$Ag, $^{89}$Rh and $^{187}$Bi for our purpose to be studied with FSU Gold. The binding energy results
are already given in Table \ref{be}.
For the first three decays, the Q-values are estimated and tabulated 
in  \cite{mt} and presented in Table \ref{life3}.
The first is known to undergo $\beta$-decay while the latter two 
may possibly decay also via proton emission.
The Q-value for  $^{187}$Bi  can be calculated from the binding energy 
difference between the parent and the daughter
and is taken to be 1.130(19) MeV.
Similar to the situation in 
$^{185}$Bi, the  life time is calculated for the decay from the $1/2^+$ state 
of the parent to the ground state of the daughter.  

The results for the first three decays are tabulated in Table \ref{life3}.
The ground state of the nuclei $^{89}$Rh, $^{93}$Ag and $^{97}$In are all 
assumed to be $9/2^+$. This follows from the systematics as well as the shell 
model picture. Only the lower limit of the  half life values of the first two 
nuclei are known and both are 1.5 $\mu$s. The half life of $^{97}$In
is taken to be 5 ms from systematics. Assuming a possible proton radioactivity 
from the ground state of the parent to the ground state of the daughter, 
the half life values have been calculated using the DDM3Y interaction. 
As the Q-values in these nuclei are not known, we have calculated 
the upper and the lower limits of the half life values corresponding to the 
two limits of Q-values also. The picture is not clear due to the very large 
uncertainties in the Q-values as well as the experimental half life values. 
However, a few general
remarks may be made. If we assume a half life value close to the lower limit
in $^{89}$Rh, we see that the proton radioactivity has a very small branching
ratio compared to $\beta$-decay. However, in $^{93}$Ag, it may be possible 
to observe proton radioactivity, indeed it may be the dominant form of decay.
The situation in $^{97}$In is more complicated. Here, even assuming the longest 
half life, the dominant decay mode is expected to be proton radioactivity. 
The experimental life time value is from systematics and may not be correct.
From the preceding discussion, it is clear that possible $^{93}$Ag and 
$^{97}$In are potential candidates for proton radioactivity at ground state.

With respect to the $1/2^+$ excited state in $^{187}$Bi, this level
is seen to have a half life of 0.290 ms\cite{Bi187} and has a branching 
ratio $>50\%$ \cite{Bi1871} for alpha decay. The calculated half life
for proton radioactivity from this state, assuming a 9\% admixture of 
$2p-2h$ state in the ground state of $^{186}$Pb, is 10.9 seconds. Thus
the branching ratio for proton radioactivity is expected to be very small compared
to the alpha decay branch.

To summarize, half life values for proton emitting spherical nuclei have been 
calculated in the WKB approximation. The microscopic proton-nucleus potential
has been obtained by folding the densities of daughter nuclei with two
microscopic NN interactions, 
DDM3Y and JLM. The densities have been obtained in the RMF
approach using mainly the force FSU Gold. Results are very similar for other 
common forces. The calculated results for  the decays from
the ground state or the low-lying excited states in almost all the  
nuclei agree well with experimental measurements. Reasons for large deviations 
in a few cases have been discussed. Results in $^{109}$I and $^{112,113}$Cs 
show that the effect of deformation is small contrary to earlier calculations.
Predictions for possible proton radioactivity have been made in two nuclei, 
$^{93}$Ag and $^{97}$In.

                                                                                
This work was carried out with financial assistance of the
Board of Research in Nuclear Sciences, Department of Atomic Energy (Sanction
No. 2005/37/7/BRNS). Discussions with Subinit Roy is gratefully acknowledged.
                                                                                

\begin{table}[h]
\begin{center}
\caption{Binding energy values of daughter nuclei in the decays studied in the present work.
Experimental values are from \cite{mt}.\label{be}}
\begin{tabular}{rlcrlc}\hline
Nucleus &\multicolumn{2}{c}{BE/A(MeV)}& 
Nucleus &\multicolumn{2}{c}{BE/A(MeV)}\\
          & Exp.&      Calc.&        
          & Exp.&      Calc.\\\hline
$^{104}$Sn &   8.3836 &  8.3507& 
$^{108}$Te &   8.3032 &  8.2575\\
$^{111}$Xe &   8.181* &  8.1386& 
$^{112}$Xe &   8.2295 &  8.1892\\
$^{144}$Er &   7.958* &  7.8844& 
$^{146}$Er &   8.013* &  7.9440\\
$^{149}$Yb &   7.929* &  7.8905& 
$^{150}$Yb &   7.964* &  7.9376\\
$^{154}$Hf &   7.918* &  7.9126& 
$^{155}$Hf &   7.928* &  7.9207\\
$^{156}$Hf &   7.9529 &  7.9365& 
$^{159}$W  &   7.866* &  7.8585\\
$^{160}$W  &   7.8930 &  7.8780& 
$^{163}$OS &   7.805* &  7.8079 \\    
$^{164}$Os &   7.8335 &  7.8313& 
$^{165}$OS &   8.842* &  7.8467\\
$^{166}$Os &   7.8864 &  7.8696& 
$^{170}$Pt &   7.8083 &  7.8354\\
$^{176}$Hg &   7.7826 &  7.8031& 
$^{184}$Pb &   7.7827 &  7.7828\\\hline
$^{88}$Ru  &   8.313* & 8.2009 &
$^{94}$Pd  &   8.283* & 8.1912  \\
$^{96}$Cd   &  8.265*  & 8.2040       &
$^{186}$Pb &   7.8053 & 7.7901  \\\hline
\end{tabular}
\end{center}
\flushleft
*Estimated value
\end{table}
\begin{table}[h]
                                                                                
\caption{
Experimental and calculated proton decay half lives (T) for spherical proton 
emitters with even neutron number. The densities have been calculated using
the force FSU Gold. Here DDM3Y and JLM indicate the calculated half life values 
obtained using the respective effective NN interactions. The last column 
presents the results of the calculation of \cite{oth}. The 
angular momentum of the proton involved is given by $l$. The experimental 
Q, $l$ and half life values are from Ref. \cite{nds}.
\label{life1}}
\begin{tabular}{lcllcclll}\hline
Nucleus & $l$ &   Q(MeV)  &\multicolumn{6}{c}{log$_{10}$T(s)}\\\cline{4-9}
        &     &      &\multicolumn{3}{c}{Exp.}        &\multicolumn{2}{c}{Present work} & Ref\cite{oth}\\\cline{4-6}
  &($\hbar$)&&& \multicolumn{2}{c}{Errors}&DDM3Y& ~~~~JLM& DDM3Y\\\hline
$^{105}$Sb  &  2 &  0.491(15)  &  ~~2.049 & -0.067 & +0.058 & ~~2.27(46) &   ~~1.69(45) & ~~1.97(46)   \\
$^{109}$I   &  2 &  0.829(3)  & $-3.987$ & -0.022 & +0.020 & $-4.03(4)$ &  $-4.01(4)$& $-4.25$\\
$^{113}$Cs  &  2 &  0.978(3)  & $-4.777 $& -0.019 & +0.018 & $-5.34(4)$ &  $-5.32(4)$ & $-5.53^\dagger$\\
$^{145}$Tm  &  5 &  1.753(10)  & $-5.409 $&-0.146 & +0.109  & $-5.20(6)$ &  $-5.10(6)$&  $ -5.14(6)$ \\
$^{147}$Tm  &  5 &  1.071(3)  &  ~~0.591 & -0.175 & +0.125 & ~~0.98(4) &   ~~1.07(4)&   ~~0.98(4) \\
$^{147}$Tm* &  2 &  1.139(5)  & $-3.444$ & -0.051 & +0.046 & $-3.26(6)$ &  $-3.27(6)$&   $-3.39(5)$ \\
$^{151}$Lu  &  5 &  1.255(3)  & $-0.896$ & -0.012 & +0.011 & $-0.65(3)$ &  $-0.55(3)$&   $-0.67(3)$ \\
$^{151}$Lu* &  2 &  1.332(10)  & $-4.796$ &-0.027 & +0.026 & $-4.72(10)$ &  $-4.73(10)$&   $-4.88(9)$ \\
$^{155}$Ta  &  5 &  1.791(10)  & $-4.921$ & -0.125 & +0.125 & $-4.67(6)$ &  $-4.57(6)$&   $-4.65(6)$ \\
$^{157}$Ta  &  0 &  0.947(7)  & $-0.523$ & -0.198 & +0.135   & $-0.21(11)$ &  $-0.23(11)$&   $-0.43(11)$ \\
$^{161}$Re  &  0 &  1.214(6)  & $-3.432$ & -0.049 & +0.045   & $-3.28(7)$ &  $-3.29(7)$&   $-3.46(7)$ \\
$^{161}$Re* &  5 &  1.338(7)  & $-0.488$ & -0.065 & +0.056 & $-0.57(7)$ &  $-0.49(7)$&   $-0.60(7)$ \\
$^{165}$Ir* &  5 &  1.733(7)  & $-3.469$ & -0.100 & +0.082 & $-3.52(5)$ &  $-3.44(5)$&   $-3.51(5)$ \\
$^{167}$Ir  &  0 &  1.086(6)  & $-0.959$ & -0.025 & +0.024 & $-1.05(8)$ &  $-1.07(8)$&   $-1.27(8)$ \\
$^{167}$Ir* &  5 &  1.261(7)  & $~~ 0.875$ & -0.127 & +0.098 & $~~~0.74(8)$ &  $~~ 0.81(8)$&   $~~0.69(8)$ \\
$^{171}$Au  &  0 &  1.469(17)  & $-4.770$ & -0.151 & +0.185 & $-4.84(15)$ &  $-4.86(15)$&   $-5.02(15)$ \\
$^{171}$Au* &  5 &  1.718(6)  & $-2.654$ & -0.060 & +0.054& $-3.03(4)$ &  $-2.96(4)$&   $-3.03(4)$ \\
$^{177}$Tl  &  0 &  1.180(20)  & $-1.174$ & -0.349 & +0.191 & $-1.17(25)$ &  $-1.20(25)$&   $-1.36(25)$ \\
$^{177}$Tl* &  5 &  1.986(10)  & $-3.347$ & -0.122 & +0.095 & $-4.52(5)$ &  $-4.46(5)$&   $-4.49(6)$ \\
$^{185}$Bi  &  0 &  1.624(16)  & $-4.229$ & -0.081 & +0.068 & $-5.33(13)$ &  $-5.36(13)$&   $-5.44(13)$ \\\hline
\end{tabular}

$^\dagger$ Calculated for a Q -value of 0.977 MeV.
\end{table}

\begin{table}[h]
                                                                                
\caption{Experimental and calculated proton decay half lives (T) for spherical 
proton emitters with odd neutron number. See caption of table \ref{life1}
for details.\label{life2}}
\begin{tabular}{lcllcclll}\hline
Nucleus & $l$ &   Q(MeV)  &\multicolumn{6}{c}{log$_{10}$T(s)}\\\cline{4-9}
        &     &      &\multicolumn{3}{c}{Exp.}        &\multicolumn{2}{c}{Present work} & Ref\cite{oth}\\\cline{4-6}
  &($\hbar$)&&& \multicolumn{2}{c}{Errors}&DDM3Y& ~~~~JLM& DDM3Y\\\hline
$^{112}$Cs  &  2 &  0.824(7) & $-3.301 $& -0.097 & +0.079    & $-2.93(11)$ &  $-2.91(11)$ & $-3.13^\dagger$\\
$^{150}$Lu  &  5 &  1.283(4) & $-1.180 $& -0.064 & +0.055    & $-0.59(4)$ &  $-0.49(4)$ & $ -0.58(4)$ \\
$^{150}$Lu* &  2 &  1.317(15) & $-4.523 $& -0.301 & +0.620    & $-4.24(15)$ &  $-4.24(15)$ & $ -4.38(15)$ \\
$^{156}$Ta  &  2 &  1.028(5) & $-0.620 $& -0.101 & +0.082    & $-0.22(7)$ &  $-0.23(7)$ & $ -0.38(7)$ \\
$^{156}$Ta* &  5 &  1.130(8) & $~~0.949$ & -0.129 & +0.100   & $~~1.66(10)$ &  $~~1.76(10) $&  $~~1.66(10)$\\
$^{160}$Re  &  2 &  1.284(6) & $-3.046$ & -0.056 & +0.075   & $-2.86(6)$ &  $-2.87(6)$ & $ -3.00(6)$ \\
$^{164}$Ir  &  5 &  1.844(9) & $-3.959$ & -0.139 & +0.190   & $-3.95(5)$ &  $-3.86(5)$ & $ -3.92(5)$ \\
$^{166}$Ir  &  2 &  1.168(8) & $-0.824$ & -0.273 & +0.166   & $-0.96(10)$ &  $-0.96(10)$ & $ -1.11(10)$ \\
$^{166}$Ir* &  5 &  1.340(8) & $-0.076$ & -0.176 & +0.125    & $~~0.22(8)$ &  $~~0.30(8)$ & $~ ~0.21(8)$ \\\hline
\end{tabular}

$^\dagger$ Calculated for a Q-value of 0.823 MeV.
\end{table}

\begin{table}[h]
                                                                                
\caption{Binding energy and proton decay half life values calculated 
using different RMF forces.\label{rmf}}
\begin{tabular}{lccrrrr}\hline
Nucleus &\multicolumn{3}{c}{BE/A(MeV) of daughter}&\multicolumn{3}{c}{log$_{10}$T(s)}\\
& FSU Gold & NL3 & NLSH & FSU Gold & NL3 & NLSH \\\hline
$^{113}$Cs & 8.1892 & 8.1868 & 8.2270 & $-5.34$ & $-5.35$ & $-5.32$\\
$^{147}$Tm & 7.9440 & 7.9657 & 7.9612 & 0.98 & 0.96 & 1.00 \\
$^{147}$Tm*&        &        &        & $-3.26$ & $-3.26$ & $-3.24$\\
$^{185}$Bi & 7.7828 & 7.7817 & 7.7903 & $-5.33$ & $-5.34$ & $-5.31$\\\hline
\end{tabular}
\end{table}

\begin{table}[h]
                                                                                
\caption{Calculated proton decay half lives (T) for $^{89}$Rh, $^{93}$Ag, 
$^{97}$In. The experimental half life in seconds including all the decays 
is denoted by $\tau$. Here I, II and III in the columns for calculated 
values refer respectively to the half lives for the mean Q-value, 
followed by the ones corresponding to its upper and lower limits.
See caption of table \ref{life1} for more details.\label{life3}}
\begin{tabular}{lccrcrc}\hline
Nucleus & $l$ &   Q(MeV)  &log$_{10}\tau$(s)&\multicolumn{3}{c}{log$_{10}$T(s)}\\
  &($\hbar$)&           &     Exp. &I &II &III \\\hline
 $^{89}$Rh & 4 & 0.700(200) & $>-5.824$ & $-0.36$ & $-3.07$ & 3.76\\
 $^{93}$Ag & 4 & 1.430(780) & $>-5.824$ & $-8.76$ & $-12.15$ & $-0.66$\\
 $^{97}$In & 4 & 1.812(780) & $-2.301$   & $-10.33$ &$-12.97$& $-5.21$ \\
\hline
\end{tabular}
\end{table}

\begin{thebibliography}{99}
\bibitem{prdecay}P.J. Woods and C.N. Davis, Annu. Rev. Nucl. Part. Sci. {\bf 47} (1997) 541.­
\bibitem{nds} A.A. Sonzogni, Nucl. Data Sheets {\bf 95} (2002) 1.
\bibitem{Buck}B. Buck, A. C. Merchant and S. M. Perez, Phys. Rev. C {\bf 45} (1992) 1688.                                                                      \bibitem{Aberg}S. Aberg, P. B. Semmes and W. Nazarewicz, Phys. Rev. C {\bf 56} (1997) 1762.
\bibitem{mfl} E. Maglione, L.S. Ferreira and R.J. Liotta,\PRL {\bf 81} (1998) 538 ; \PR C {\bf 59} (1999) R589. 
\bibitem{oth} D. N. Basu, P. Roy Chowdhury and C. Samanta, \PR {\bf 72} (2005)  
051601(R) ; P. Roy Chowdhury, C. Samanta and  D. N. Basu, Arxiv 
Nucl-Th (2005) 0511090.
\bibitem{RMF1}See {\em eg.} P. Ring, Prog. Part. Nucl. Phys. {\bf 37} (1996) 193. 

\bibitem{rhb} G. A. Lalazissis, D. Vretenar and P. Ring, \NP          
{\bf A650} (1999) 133 ; {\bf A 719} (2003) C209.
\bibitem{prl} J. Piekarewicz J and B.G. Todd-Rutel, \PRL {\bf 95} (2005) 122501. 

\bibitem{NL3}  G.A. Lalazissis, J. K\"{o}nig and P. Ring, \PR C{\bf 55} (1997) 540.
\bibitem{NLSH}M.M. Sharma, M.A. Nagarajan and P. Ring, \PL {\bf B312} (1993) 377.
\bibitem{ddm3y1} A. M. Kobos, B. A. Brown, R. Lindsay and G. R. Satchler,
Nucl. Phys. {\bf A425} (1984) 205. 
\bibitem{ddm3y2} A. K. Chaudhuri, Nucl. Phys. {\bf A449} (1986) 243 ; {\bf A459} (1986) 41. 
\bibitem{ddm3y3} D.N. Basu, \jpg {\bf 30} (2004) B7.
\bibitem{jlm} J.P. Jeukenne, A. Lejeune and C.Mahaux, \PR C {\bf 14} (1974) 1391.
\bibitem{jlm1} E. Bauge, J.P. Delaroche and M. Girod, \PR C{\bf 63} (2001) 024607.
\bibitem{CBe} G. Gangopadhyay and S. Roy, \jpg {\bf 31} (2005) 1111.
\bibitem{ev} D. N. Poenaru, W. Greiner, M. Ivascu, D. Mazilu and I. H. 
Plonski, Z. Phys. A {\bf 325} (1986) 435.
\bibitem{mt} G. Audi G, A.H. Wapstra and C. Thibault,  \NP {\bf A729} (2003) 337.
\bibitem{Liu} Z. Liu \etal, \PR C{\bf 72} (2005) 047301.
\bibitem{Bi185}C.N. Davids \etal, \PRL {\bf 76} (1996) 592.
\bibitem{Pb} J. Wauters \etal, \PR C {\bf 50}, 2768 (1994).
\bibitem{sjnp} V.P. Bugrov and S.G. Kadmenskyi, Sov. J. Nucl. Phys. {\bf 49} (1989) 967.
\bibitem{Bi187}J.C. Batchhelder \etal,  Eur. Phys. J. A {\bf 5} (1999) 49.
\bibitem{Bi1871}E. Coenen, K. Deneffe, M. Huyse, P. Van Duppen and J.L. 
Wood, \PRL {\bf 54} (1985) 1783.

\end{thebibliography}
\end{document}